# "Task-relevant autoencoding" enhances machine learning for human neuroscience

**Authors:** Seyedmehdi Orouji[1], Vincent Taschereau-Dumouchel[2-3], Aurelio Cortese[4], Brian Odegaard[5], Cody Cushing[6], Mouslim Cherkaoui[6], Mitsuo Kawato[4], Hakwan Lau[7], & Megan A. K. Peters[1,8]

1 Department of Cognitive Sciences, University of California, Irvine, Irvine, California, USA 92697
2 Department of Psychiatry and Addictology, Université de Montréal, Montreal, Canada, H3C 3J7.
3 Centre de recherche de l'institut universitaire en santé mentale de Montréal, Montréal, Canada.
4 ATR Computational Neuroscience Laboratories, Kyoto, Japan 619-0288
5 Department of Psychology, University of Florida, Gainesville, FL USA 32603
6 Department of Psychology, University of California Los Angeles, Los Angeles, 90095, USA
7 RIKEN Center for Brain Science, Tokyo, Japan
8 Center for the Neurobiology of Learning and Memory, University of California, Irvine, Irvine, California, USA 92697

**Correspondence should be directed to:**

Seyedmehdi Orouji
Department of Cognitive Sciences
2201 Social & Behavioral Sciences Gateway
University of California, Irvine
Irvine, CA 92697
sorouji@uci.edu

Megan A. K. Peters
Department of Cognitive Sciences
2201 Social & Behavioral Sciences Gateway
University of California, Irvine
Irvine, CA 92697
megan.peters@uci.edu

## Abstract

In human neuroscience, machine learning can help reveal lower-dimensional neural representations relevant to subjects' behavior. However, state-of-the-art models typically require large datasets to train, so are prone to overfitting on human neuroimaging data that often possess few samples but many input dimensions. Here, we capitalized on the fact that the features we seek in human neuroscience are precisely those relevant to subjects' behavior. We thus developed a Task-Relevant Autoencoder via Classifier Enhancement (TRACE), and tested its ability to extract behaviorally-relevant, separable representations compared to a standard autoencoder, a variational autoencoder, and principal component analysis for two severely truncated machine learning datasets. We then evaluated all models on fMRI data from 59 subjects who observed animals and objects. TRACE outperformed all models nearly unilaterally, showing up to 12% increased classification accuracy and up to 56% improvement in discovering "cleaner", task-relevant representations. These results showcase TRACE's potential for a wide variety of data related to human behavior.

**Keywords:** human neuroscience, machine learning, dimensionality reduction, task-relevant representation, fMRI, MVPA, autoencoder

# 1. Introduction

In studying the human brain and human behavior, we often use machine learning methods to home in on the (ideally lower-dimensional[1–4]) representations contained in multivariate, feature-rich datasets. These data typically contain noisy, irrelevant signals[19–21] that we would like to filter out using methods such as multivariate decoders[5–8], various types of autoencoders, generative adversarial networks like InfoGAN[9], or even principal components analysis (PCA)[10–12]. However, state-of-the-art machine learning methods typically require very large datasets to train while data for individual human subjects collected with methods such as functional magnetic resonance imaging (fMRI)[13–15] are often severely limited in sample size[16,17] (i.e., have very few training exemplars compared to the dimension of data). Consequently, these methods are susceptible to overfitting on such neuroimaging data, reducing their predictive power and utility[18–20]. What's more, parametric methods (such as PCA), which may better avoid the need for large training sets, by definition require rigid assumptions regarding the nature of the dimensionality reduction process and thus are limited *a priori* to insights consistent with these parametric assumptions. Thus, we are in need of a nonparametric method that can reveal the *low-dimensional, task-relevant* representations in a given brain region using *exemplar-poor but input-dimension-rich* datasets.

Here, we sought to capitalize on a unique property of many human neuroimaging datasets, which is that the features we wish to identify can be conceptualized based on whether they are relevant for the subject's behavior.

We drew inspiration from previous successes with classifier-enhanced autoencoders[21–24] to develop the Task-Relevant Autoencoder via Classifier Enhancement (TRACE) model. TRACE's architecture is purposely simple to limit overfitting to small datasets, consisting of a fully-connected autoencoder with only one hidden layer on each of the encoding and decoding arms and a logistic regression classifier attached to the bottleneck layer (**Figure 1**).

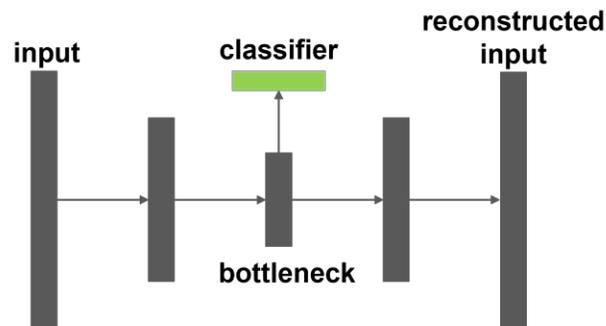

**Figure 1.** A cartoon representation of the TRACE network architecture. Each gray rectangle represents a layer of the autoencoder, consisting of fully connected units. The input layer is connected to the bottleneck layer via one hidden encoding layer, and again to the reconstruction layer via one hidden decoding layer. A classifier is attached to the bottleneck and contributes to the objective optimization function.

We developed four quantitative metrics to assess TRACE's performance at different bottleneck dimensionalities (compression levels), and then comprehensively benchmarked TRACE under conditions of severe data sparsity using the MNIST[25] and Fashion MNIST[26] datasets, two of the most popular machine learning datasets. We then applied TRACE to a neuroimaging (fMRI) dataset of subjects who viewed and categorized animals and objects while blood oxygen level dependent (BOLD) signal was collected from ventral temporal cortex (VTC) in a single, 1-hour session. By constraining the dimensionality reduction process to specifically prioritize features that were relevant to the participants' behavioral task, we show that TRACE can extract both quantitatively and qualitatively 'cleaner' representations at both reduced dimensions and in the original input dimensionality, showing up to threefold improvement in decoding accuracy and separation of class-specific patterns. These results demonstrate our method can distill highly separable, low dimensional neural representations even with sparse and noisy data. TRACE may thus show promise on a broad variety of behaviorally-relevant neuroimaging datasets.

## 2. Results

We quantified the performance of the Task-Relevant Autoencoder via Classifier Enhancement (TRACE) model against that of a standard autoencoder (AE), a Variational Autoencoder (VAE), and using principal component analysis (PCA) via (1) *reconstruction fidelity,* (2) *reconstruction classifier accuracy*, (3) *bottleneck classifier accuracy,* and (4) *reconstruction class specificity* (see **Methods Section 4.4**) ("class" here refers to the class of the input image, e.g. "9" or "shoe" or "cat"). We assessed these metrics as a function of different bottleneck dimensionalities (i.e., compression levels), first on the MNIST and Fashion MNIST datasets under increasing data sparsity and then on a previously-collected fMRI dataset of ventral temporal cortex (VTC) (i.e., voxel activations while 59 human subjects viewed 40 classes of animals and objects). We also performed additional investigation at each dataset's 'optimal' bottleneck dimensionality (where reconstruction class specificity is maximized) to characterize each model's behavior.

### 2.1 Benchmarking TRACE's advantages, including under increasing data sparsity

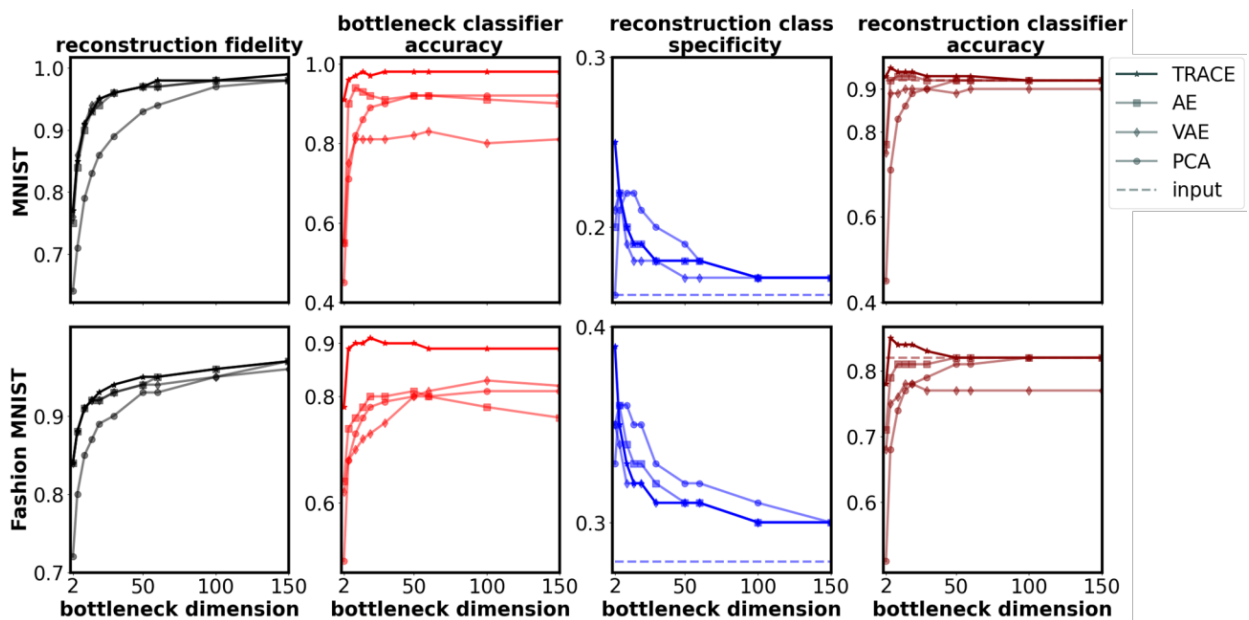

**Figure 2.** Quantitative comparison between TRACE and other models (AE, VAE, and PCA) on the four outcome metrics, for the two benchmark datasets (MNIST & Fashion MNIST) for bottleneck dimensionalities between 2 and 150. All metrics show superiority of TRACE over other models (except higher dimensionalities in reconstruction class specificity). TRACE is shown by the darker line while other models are shown by lighter lines. The black, red, blue, and dark red lines show the reconstruction fidelity, bottleneck classifier accuracy, reconstruction class specificity, and reconstruction classifier accuracy, respectively (see **Methods**). The dashed blue and dark red lines show the input class specificity and input classifier accuracy, respectively. Outcome metrics for all bottleneck dimensionalities tested (2-784) are shown in **Figure S3**; locations of peaks for all four metrics are shown in **Table S1.** The chance levels of bottleneck and reconstruction classifier accuracy are both 10% (not shown in the plot).

We first examined *reconstruction fidelity* (black, **Figure 2**), i.e. the mean Pearson correlation of the inputs and corresponding reconstructions. High reconstruction fidelity assures us that the discovered features in the bottleneck provide a reasonable representation of this high-dimensional information – i.e., that the autoencoder portion of the models can be successfully trained. Notably, TRACE's reconstruction fidelity performed in a similar fashion despite the fact that the contribution of the reconstruction part of the loss function (mean square error; MSE) for TRACE was smaller than for AE and VAE (i.e., the objective function in TRACE is the sum of reconstruction loss ($L_R$) and classification loss functions ($L_{CE}$); see **Methods Section 4.4.1**).

Next, we examined *bottleneck classifier accuracy* (bright red, **Figure 2**), i.e. the accuracy of a separate classifier trained with bottleneck features as input after the training of all models. Bottleneck classifier accuracy was much higher for TRACE than for other models even at very low bottleneck dimensionalities. As bottleneck dimensionality grew, this metric asymptotically equalizes to at least ~10% better than all other models in the MNIST and Fashion MNIST datasets. Notably, though, in both datasets, at all bottleneck dimensionalities tested, TRACE bottleneck classifier accuracy is *always* higher than that for other models. Although by attaching a cross entropy loss function to the bottleneck of the network one can expect to naturally discover features that increase the classification accuracy, crucially this achievement is gained without losing the ability to reconstruct the input in the decoder part of the network. In other words, the lower dimensional representations learned by TRACE are not only more suitable for classification purposes but also can be used just as effectively to reconstruct the input.

The third metric we examined was *reconstruction class specificity* (blue, **Figure 2**), i.e. the average within-class correlation of the reconstructed inputs minus the average between-class correlation. This metric quantifies the degree of separation between class clusters in reconstruction feature space as a measure of reconstruction representations' categorical 'purity'. Reconstruction class specificity peaks at bottleneck dimensionality d=2 for TRACE for both MNIST and Fashion MNIST. As with the other metrics, TRACE outperformed other models at optimal bottleneck dimensionality d=2.

Fourth, we examined *reconstruction classifier accuracy* (dark red, **Figure 2**), i.e. the accuracy of a separate logistic classifier trained to discriminate classes using reconstructed data. Reconstruction classifier accuracy quantifies the task-relevance of the information extracted through the compression process, and also provides a direct benchmark against which to compare to the noisiness of representations in the original input space (see below). Reconstruction classifier accuracy for both MNIST and Fashion MNIST peaked at bottleneck dimensionality d=5 for TRACE, and was consistently higher for TRACE over other models.

Interestingly, that this metric peaks at higher bottleneck dimensionalities than reconstruction class specificity suggests that the performance of a classifier trained on these high-dimensional reconstructions may not meaningfully reflect the maximum compression that TRACE can achieve without loss of overall performance.

A final – and critical – test of TRACE would examine its ability to not only distill task-relevant information into low-dimensional representations but also 'push' such distilled insights back into the native space of the input. This would be especially important if one wished to use TRACE to de-noise fMRI data to discover multivoxel patterns representing a target concept or category to be used with noninvasive intervention strategies such as decoded neurofeedback (DecNef)[27–30] (or to simply investigate those activity patterns in native space). Although iterative sparse logistic regression and support vector machine classification have been demonstrated as successful at identifying such patterns when trained on the native input data[27,31,32], we wanted to see whether TRACE would be able to denoise the data such that an even cleaner target pattern would become discoverable. Specifically, if TRACE is successful at actively removing task-irrelevant noise rather than simply passively averaging across it (as is done with a standard category-based logistic regression) or removing it through iterative sparsity approaches (iterative sparse logistic regression), then we should observe two patterns. First, reconstruction classifier accuracy should approach or exceed classification accuracy of an identical logistic regression classifier trained on the native inputs. Second, reconstruction class specificity should behave similarly, approaching and then exceeding input class specificity. This behavior makes reconstruction class specificity an ideal metric for defining the 'optimal bottleneck dimensionality' if one's goal is to optimally distill representations in native space.

To evaluate this behavior, we (a) trained an additional logistic regression classifier on each of the datasets to classify the native input, and (b) computed class specificity directly from the raw input data for all three datasets. We then compared the outcomes to the reconstruction classifier accuracy and reconstruction class specificity computed as a function of bottleneck dimensionality.

Results revealed that, for MNIST, reconstruction classifier accuracy (solid dark red, **Figure 2**) exceeded input classifier accuracy (dashed dark red line) immediately (at d=2) for TRACE but not until d=10 for AE; it never exceeded the input for other models. For Fashion MNIST, this occurred at d=5 for TRACE and only at much higher dimensionality – if at all – for the other models tested. These results show that TRACE provides not only superior compression but also superior denoising even in comparison to the direct inputs. TRACE's denoising capability can be particularly useful in DecNef[27–29,33–39] studies as it can minimize the task-irrelevant information of exemplars even in the anatomical and functional brain space.

Results for reconstruction class specificity followed a different pattern, but still favored TRACE: reconstruction class specificity (solid blue line, **Figure 2**) exceeded input class specificity (dashed blue line) at most bottleneck dimensionalities for all models, but was higher for TRACE at the optimal bottleneck dimensionality (d=2). These results show that TRACE can provide a powerful method for not only distilling low-dimensional representations, but also in pushing those cleaner representations back into the structure and dimensionality of the raw input space. That is, a structurally identical logistic classifier with the same number of parameters can exhibit better performance using the reconstructed inputs than using the original inputs themselves.

Note that conducting statistical tests of the results from **Figure 2** is not feasible since the results reported here come from the training of models on the entire dataset at each dimensionality of the bottleneck.

### 2.1.1 Comprehensive comparison across metrics as a function of increasing data sparsity

We next sought to select a single bottleneck dimensionality for TRACE to explore its benefits over AE, VAE, and PCA under increasing data sparsity. For this purpose, we selected the maximal value of reconstruction class specificity because this metric provides the best balance between task-relevant information extraction and compression, both for analyzing low-dimensional representations and patterns in the original input dimensionality (e.g., for use with real-time DecNef[27–29,33–39]).

Reconstruction class specificity peaked at d=2 for both MNIST and Fashion MNIST, so we can first examine TRACE's superiority at this dimensionality when maximal data is available (n = 60,000 training samples for both datasets). Since the goal is to compress information as much as possible without losing information, we chose d=2 to conduct the rest of the analysis given TRACE peaks at a bottleneck dimensionality lower than AE or other models (i.e., d=2). Additionally, at TRACE's peak (d=2), TRACE shows superior performance compared to other models' performance at dimensionalities where those other models peak (e.g., TRACE's reconstruction class specificity is higher at TRACE's peak [d=2] than AE's reconstruction class specificity is at AE's peak [d=5]). Here, we see that TRACE's superior extraction of task-relevant information comes at no loss in reconstruction fidelity over AE (**Figure 2**; **Table S2**). Further explorations, described below, were therefore done at bottleneck dimensionality d=2.

To examine how TRACE fared versus the other models under increasing data sparsity, we trained each model after removing 10, 30, 50, 70, 90, 95, and 98 percent of the training data. Training examples at each level of sparsity for all models remained the same. We then used the conventional 10,000 held-out test set on the trained models and calculated all four metrics for all levels of data sparsity.

TRACE was much more robust to increasing data sparsity than other models (**Figure 3**). Specifically, TRACE's performance was much better even when only 2% of the data (1200 samples) remained available for training. (The test set remained fixed at the same 10,000 standard test set used for these datasets.) We note that the fMRI dataset we use below has a similar samples-to-input-dimensions ratio as the 98% truncated MNIST and Fashion MNIST datasets (~1.6 for MNIST and Fashion MNIST, and ~1.5 for this fMRI dataset). At this level of data reduction (i.e., 98% truncation) and bottleneck dimensionality d=2, we performed 50 jack-knife replications to select 2 percent of exemplars in MNIST and Fashion MNIST for training, and reported the mean values (calculated within the standard test set) of the 50 independent training sets for all metrics. As shown in **Figure 3**, TRACE continued to demonstrate superior performance even at the most extreme level of data truncation (i.e., 98% truncation). TRACE nearly uniformly swept other models across all performance metrics. To confirm TRACE's superior performance, we ran four one-way repeated measures ANOVAs at 98% truncation – one for each of the outcome metrics – with factor model (4 levels). We then followed each omnibus ANOVA with planned contrasts comparing TRACE to each other model in a pairwise fashion. This analysis revealed a main effect of model for all four outcome metrics, and that TRACE was statistically superior to all other models in all 12 pairwise comparisons; see **Table S3** for all statistics).

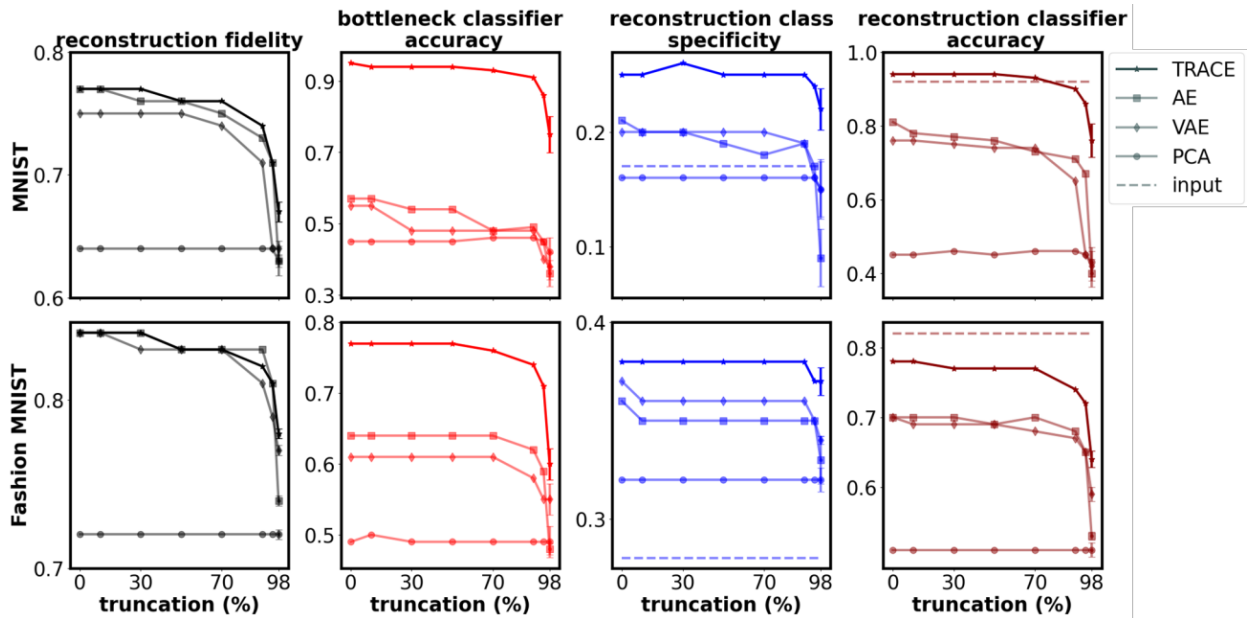

**Figure 3.** Performance of TRACE and other models as a function of sample size for the optimal bottleneck dimension of d=2. At 98% truncation level, we used 50 independent jack-knife resamplings to truncate 98 percent of exemplars and reported the means and standard deviations of the metrics (calculated on the standard test set) for MNIST and Fashion MNIST. Error bars show the standard deviation of results across the 50 jack-knife resamplings at 98% data truncation. Small variations in the metrics are likely due to random initialization of weights and use of GPUs in fitting the models.

At maximal data reduction (98% truncation) and bottleneck dimensionality d=2, we then performed additional explorations of both bottleneck representations and reconstructions. First, we visualized bottleneck representations by plotting the activities of the two bottleneck features against each other for each of the 10 classes in each dataset for TRACE versus the other models (**Figure 4**). The results are striking: TRACE showed superior task-relevant representations especially for MNIST, i.e. a clear qualitative advantage in clustering performance showing distinct clusters for different classes in stark contrast to the other models' class clusters, which are heavily overlapping. Although this difference in clustering ability was less apparent for Fashion MNIST, TRACE's clusters do appear visually more tightly bound.

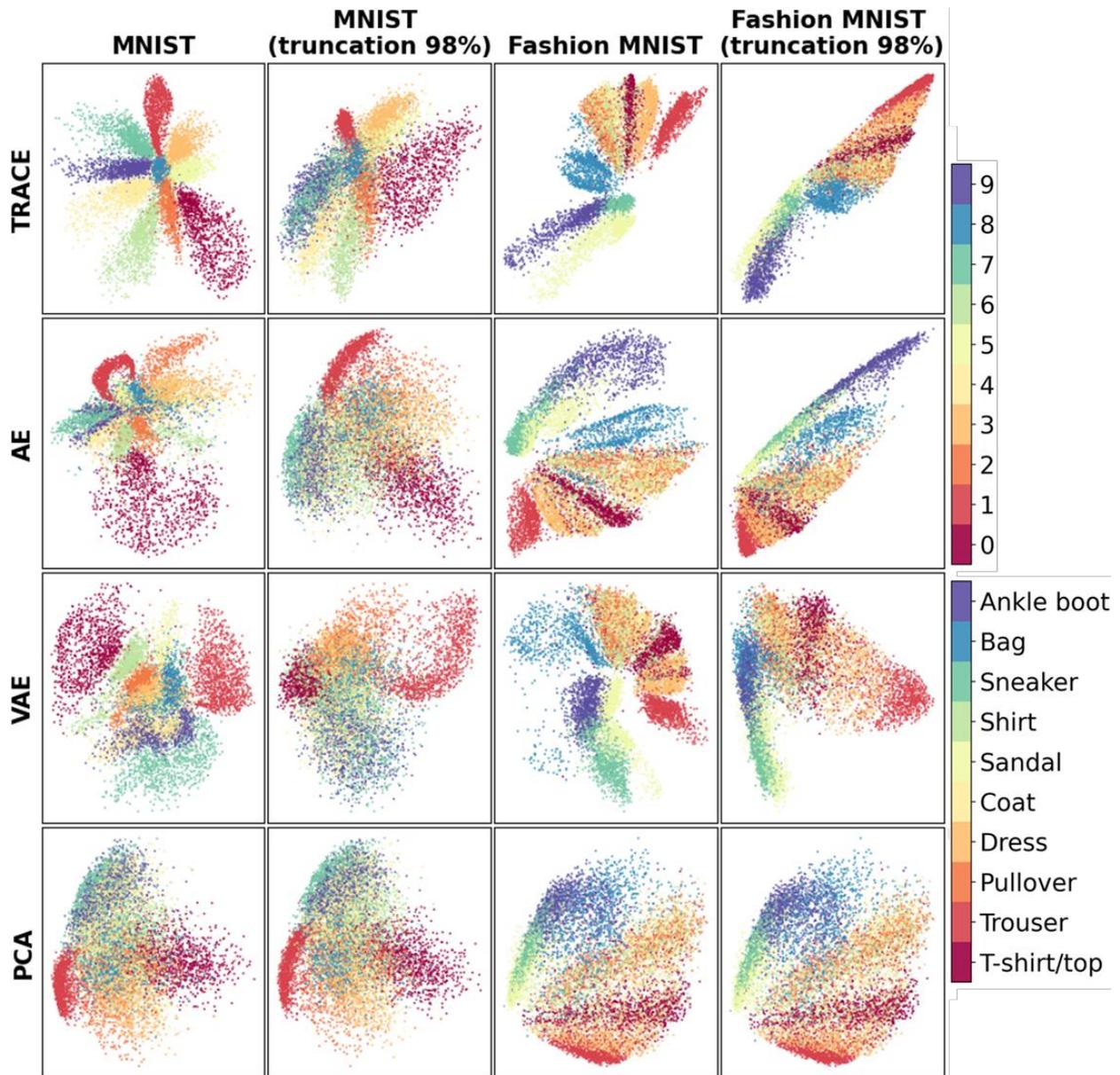

**Figure 4.** Visualization of bottleneck features for MNIST and Fashion MNIST datasets using TRACE, AE, VAE, and PCA. When trained on the full dataset, TRACE shows clear superiority in creating distinctive clusters in the bottleneck for different classes for MNIST dataset in comparison to other models The distinction is less clear but still apparent in the Fashion MNIST dataset. This pattern persists even at the 98% truncation level (trained on only 2% of the data), again showing the robustness of TRACE.

We next turned to examining the reconstructions (still at bottleneck d=2). We first examined the MNIST reconstructions for several different exemplars of the same categories (e.g., several different "3" and "6" exemplars). TRACE's superiority is clear to the naked eye: the reconstructions of particular "3" and "6" exemplars from TRACE are much more "three-like" and "six-like" than reconstructions from other models especially at the 98% truncation level (**Figure 5**). (Recall that this qualitative superiority does not come at any quantitative cost to the reconstructions). Similar findings held for Fashion MNIST (e.g., sandal and shirt), although the visual result is less striking. These patterns held even when only 2% of the data was available for training.

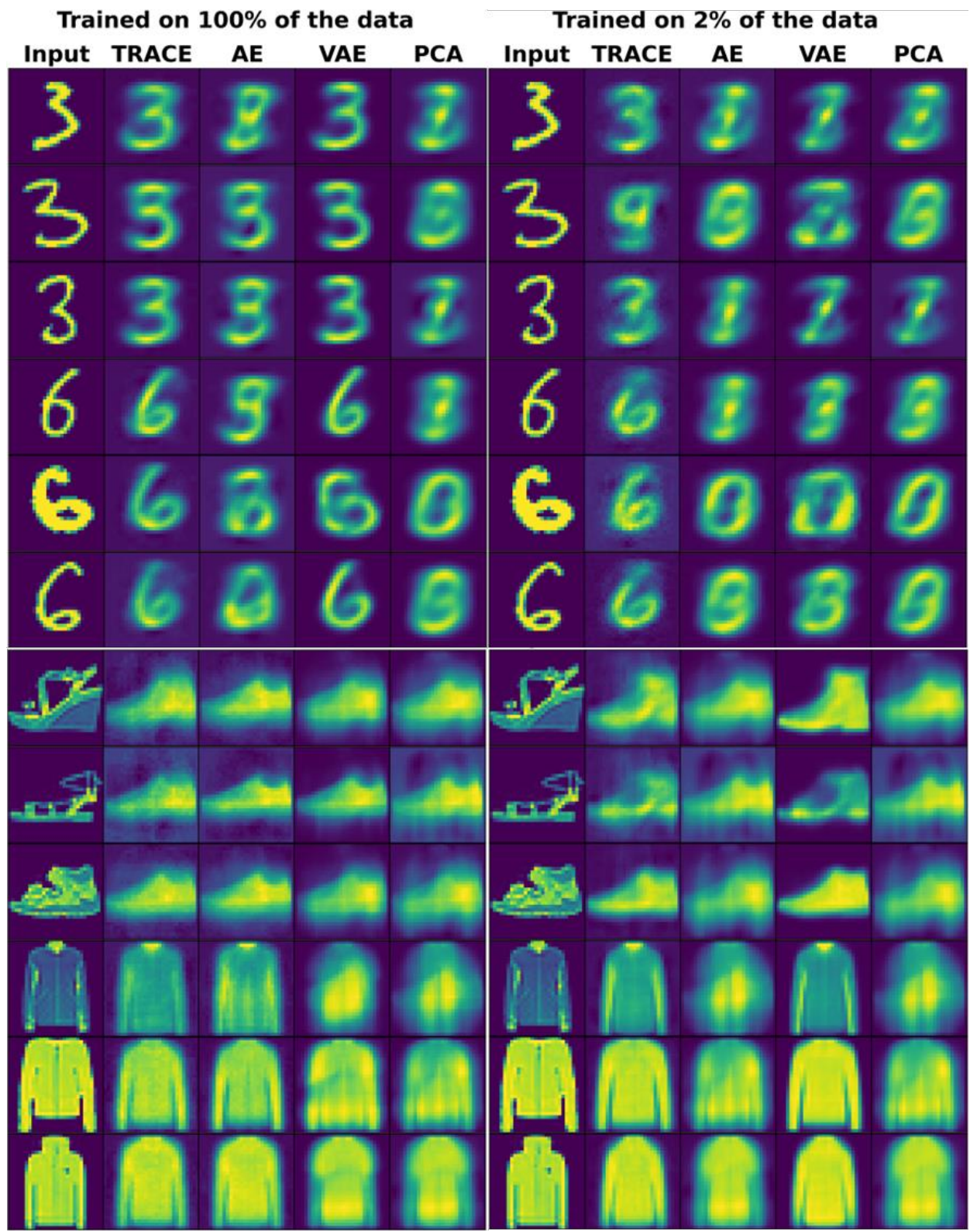

**Figure 5.** Visualization of reconstructions for MNIST and Fashion MNIST datasets using TRACE, AE, VAE, and PCA. The reconstruction of three representative instances of numbers "three" and "six" in MNIST dataset and three instances of classes "sandal" and "shirt" in the fashion MNIST dataset when there are two features in the bottleneck shows the same pattern. TRACE shows a more clear and *canonical* reconstruction of the inputs across several exemplars from the same category.

We next wanted to quantitatively investigate the distributions of within-class versus between-class clusters, both in the bottleneck and the reconstructions. This approach will facilitate evaluation of the fMRI dataset since visual inspection in fMRI data is not possible in the same sense as for MNIST and Fashion MNIST given that optimal bottleneck dimensionality is larger than 2 (**see Results Section 2.2**). We computed the effect size (Cohen's d) separating clusters in both the bottleneck and reconstructions using pairwise within- versus between-class Euclidean distances. Whether trained on all of the data or 98% truncated, Cohen's d was always larger for TRACE than for other models (**Table 1**).

|  |  | MNIST | Fashion MNIST | MNIST (98% truncation) | Fashion MNIST (98% truncation) |
|---|---|---|---|---|---|
| **Bottleneck** | **TRACE** | 1.63 ± 0.21 | 1.65 ± 0.5 | 1.36 ± 0.41 | 1.45 ± 0.35 |
|  | **AE** | 1.1 ± 0.39 | 1.5 ± 0.43 | 1.01 ± 0.41 | 1.26 ± 0.44 |
|  | **VAE** | 1.32 ± 0.55 | 1.61 ± 0.51 | 0.8 ± 0.31 | 1.48 ± 0.44 |
|  | **PCA** | 1.06 ± 0.54 | 1.25 ± 0.59 | 1.06 ± 0.54 | 1.25 ± 0.59 |
| **Reconstruction** | **TRACE** | 1.58 ± 0.21 | 1.6 ± 0.69 | 1.51 ± 0.63 | 1.54 ± 0.68 |
|  | **AE** | 1.2 ± 0.29 | 1.48 ± 0.62 | 1.02 ± 0.44 | 1.41 ± 0.66 |
|  | **VAE** | 1.27 ± 0.28 | 1.42 ± 0.61 | 0.68 ± 0.38 | 1.37 ± 0.65 |
|  | **PCA** | 1.06 ± 0.54 | 1.25 ± 0.59 | 1.06 ± 0.54 | 1.25 ± 0.59 |

**Table 1.** Cohen's d measures of effect size comparing within-class versus between-class Euclidean distances in the bottleneck and reconstructions for TRACE, AE, VAE, and PCA.

## 2.2 TRACE's performance on a real fMRI dataset

Given TRACE's apparent superiority over AE, VAE, and PCA even under extreme data sparsity, we next sought to evaluate TRACE using a real-world fMRI dataset, since ultimately our goal is to learn about neural representations. Thus, we used the same metrics as we used to evaluate TRACE on MNIST and Fashion MNIST on an fMRI dataset consisting of 59 individuals who each viewed 3600 exemplars of 40 classes of animals and objects (90 exemplars per class) while BOLD signal from ventral temporal cortex (VTC) was obtained. The number of voxels in VTC for each individual was different; however, the average of voxels for the 59 subjects was $2382 \pm 303$.

Excitingly, the fMRI dataset showed the same patterns in our four quantitative metrics as the MNIST and Fashion MNIST datasets almost across the board. First, reconstruction fidelity was actually slightly higher for AE over TRACE and VAE at higher dimensions, although this is likely due to the fact that reconstructing the input is the only objective of the AE network; however, note that the numerical difference between TRACE and AE is very small, and that both are outperforming VAE. PCA also showed higher reconstruction fidelity than all other models starting around d=500, which is also expected since as the number of principal components increases, the PCA model can explain the variance of the input data almost perfectly.

Reconstruction classifier accuracy asymptoted at bottleneck dimensionality around d=250 for all models, but again TRACE showed higher reconstruction classifier accuracy than AE, VAE, and PCA at all bottleneck dimensionalities tested. TRACE also showed higher bottleneck classifier accuracy at all bottleneck dimensionalities in comparison to other models.

TRACE outperformed other models in reconstruction class specificity as well, showing that even in the native space of the input – i.e., voxel patterns of activity in ventral temporal cortex – TRACE not only successfully distills lower-dimensional representations of task-relevant data, but also faithfully projects them back into original, high-dimensional voxel space. Reconstruction class specificity peaked at bottleneck dimensionality d=30, and then fell again. The same was not true for other models, for which reconstruction class specificity rose but then asymptoted. Crucially, though, reconstruction class specificity was also always higher for TRACE than for other models, much exceeding input class specificity (**Figure 6**, solid and dashed blue lines, respectively). This capacity to distill a task-relevant, low-dimensional representation and put it back in brain space could potentially have great value for studies in which such multivoxel patterns are the target of DecNef[27–30] or other investigations which require anatomically-related representations. We discuss this possibility in greater detail in the **Discussion**, below.

Finally, TRACE's reconstruction classifier accuracy even surpassed the input classifier accuracy for bottleneck dimensionalities higher than d=60 (dashed dark red line, **Figure 6**) which again suggests that the reconstructed version in the original input space contains more task-relevant information.

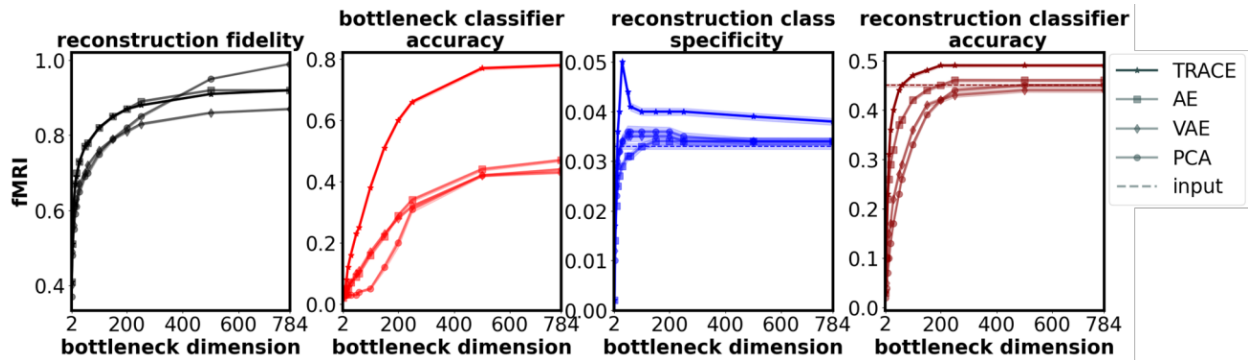

**Figure 6.** Comparison between quantitative metrics for TRACE and other models for fMRI dataset (n=59). TRACE shows superior performance in three out of four metrics (excluding reconstruction fidelity and only for d>250).

### 2.2.1 Exploration at optimal bottleneck dimensionality for fMRI data

As mentioned above, the maximal value for reconstruction class specificity was found at d=2 for the MNIST and Fashion MNIST datasets. For the fMRI dataset, we found that reconstruction class specificity peaked at d=30, so we proceeded with a parallel analysis to that done above at this dimensionality.

Crucially, at d=30, TRACE's performance on the fMRI dataset mimicked its exemplary performance on the MNIST and Fashion MNIST datasets with the exception of reconstruction fidelity, which was only slightly smaller for TRACE than for AE at dimensionalities of d>250 (and PCA at higher dimensions [i.e., d>500]) (**Figure 6**). To quantify this superiority, we performed a one-way repeated measures ANOVA for each outcome metric with factor model (4 levels), followed by planned pairwise contrasts comparing TRACE to every other model. Results revealed significant main effects of model for all four outcome metrics, and that TRACE outperformed the other models in 11 of these planned comparisons (with the exception of reconstruction fidelity between TRACE and AE; see **Table S4** for statistics).

Ultimately, as our goal is to learn about representations in human VTC, we also might want to visualize clusters for the 40 classes of the fMRI dataset. However, unlike for MNIST and Fashion MNIST where optimal bottleneck dimensionality was d=2, for the fMRI dataset we found the optimal bottleneck dimensionality at d=30. Therefore, we cannot easily visualize the class clusters in a scatterplot, and performing further dimensionality reduction for the sake of visualization would be inappropriate since assumptions of whichever dimensionality reduction technique we chose would impact the visualizations. Instead, we can use the same Cohen's d approach, described above, to characterize the tightness of the class clusters even in higher dimensionalities. The average effect size separating within- and between-class Euclidean distances across all 59 subjects was 0.38 (±0.09) for TRACE, 0.12 (±0.03) for AE, 0.11 (±0.02) for VAE, and 0.08 (±0.02) for PCA again showing TRACE's superiority.

As a final evaluation of TRACE's ability to filter out task-irrelevant information, we calculated the within- versus between-class Euclidean distance Cohen's d in the reconstructions. Pushing the distilled representations back into input space is particularly exciting for the use of TRACE with fMRI data if one wishes to discover a particular target pattern for further anatomical analysis, or for use with real-time neuroimaging (e.g., DecNef). However, visually examining fMRI reconstructions would not provide particularly useful information about the 'cleanliness' of the reconstruction, as the patterns are not visually meaningful to begin with, so we must again rely on a quantitative comparison. The average Cohen's d here again showed TRACE's superiority, with mean Cohen's d of 0.14 (±0.02) across subjects for TRACE, 0.09 (±0.02) for AE, 0.08 (±0.02) for VAE, and 0.08 (±0.02) for PCA. In other words, TRACE was able to reduce task-irrelevant information and thus extract a 'cleaner' representation, even in the reconstructions.

## 3. Discussion

### 3.1 Summary of findings

Most dimensionality-reduction approaches do not have a specific mechanism to ensure that the lower dimensional representations they reveal are particularly relevant to the question an experimenter wishes to answer. Further, many state-of-the-art deep learning models are of limited utility for discovering and characterizing meaningful representations in input-dimension-rich but exemplar-poor datasets, as they tend to overfit[18–20]. Together, these facts make discovering neural representations in within-subject fMRI datasets – which also often contain a high degree of noise and task-irrelevant information – extremely challenging[13–15]. Further, to address these issues we proposed the Autoencoder with Classifier Enhancement (TRACE) model: a simple autoencoder with a classifier attached to the bottleneck. The classifier forces the model to learn not just lower dimensional representations of the data, but those that are also task-relevant. To quantify TRACE's superiority over a standard autoencoder (AE), a variational autoencoder (VAE), and principal components analysis (PCA), we used four metrics (see **Methods Section 4.4**): 1. reconstruction fidelity; 2. bottleneck classifier accuracy; 3. reconstruction class specificity; and 4. reconstruction classifier accuracy.

TRACE outperformed all other models in all metrics, with the exception of reconstruction fidelity (sometimes). Moreover, at the 'optimal' bottleneck dimensionality, TRACE's superior capacity for extracting task-relevant information is evident in both the bottleneck and reconstruction, and TRACE's reconstructions can even outperform the inputs on a measure of task-relevant behavior (reconstruction class specificity). TRACE's advantage over other models appears due to TRACE's capacity to minimize task-irrelevant, idiosyncratic information unique to a particular sample. This is evident in the one occasional exception to TRACE's sweeping superiority: reconstruction fidelity for the fMRI dataset. However, this seeming underperformance – especially in the fMRI dataset – is actually a strength: AE tried "too hard" to encode idiosyncratic details of a particular exemplar in the bottleneck, when some of these details are merely noise for the task that the observer is performing. Thus, precise reconstruction of noisy data may not be suitable.

Critically, all of these behaviors were maintained by TRACE even under extreme data truncation for the MNIST and Fashion MNIST datasets, and carried over into a real-world fMRI dataset. These results suggest that TRACE can extract lower-dimensional representations of data for both reconstruction and classification purposes and can do so even when there is a highly undesirable balance of input-dimensions versus samples. We speculate that the better performance under the scarcity of sample size is due to adding additional label information to the bottleneck which acts as an auxiliary function to help the network to learn the general pattern in the face of scarcity of sample size. Since this scarcity of sample size is typical in fMRI data, the superior performance of TRACE suggests the strong promise of this approach for both fMRI datasets and for other biological-scale data with many more input-dimensions than samples.

## 3.2 Relation to previous work

TRACE is not the only model which can accomplish dimensionality reduction, but one of many techniques. So, is TRACE really necessary? Why would principal components analysis (PCA)[10–12] not suffice? PCA focuses on creating new features that can best explain the variance in data – including the noise and task-irrelevant information, which we know to be problematic especially in fMRI data[14,40–44] – and thus lacks explicit mechanisms to ensure the discovered lower representations contain task-relevant information. Additionally, we also note that PCA-based methods are not assumption-free (that is, they are parametric); these assumptions about the functional form of the dimensionality reduction limit the discovered features to adhering to those assumptions.

Our approach builds on previous successes with classifier-enhanced autoencoders[21–24] to extract task-relevant representations in non-biological datasets such as linguistic datasets, standard computer vision object datasets, and fault diagnosis applications. However, TRACE goes beyond these previous successes by explicitly demonstrating with otherwise matched architecture (TRACE vs AE) that the simple addition of a classifier can improve extraction of task-relevant latent representations *even under extreme data paucity*. This demonstration is especially important for the types of data used in cognitive neuroscience, which are often sample-poor. We also demonstrate that TRACE can improve reconstruction classifier accuracy and reconstruction class specificity such that it exceeds even input-level for these metrics, which could be a boon for real-time decoded neurofeedback (DecNef[27,28,37]). We will discuss these implications in more detail in Implications and Future Directions, below.

Other techniques have been developed including nonparametric techniques beyond the fully-connected AE and VAE[45,46] used here[47,48]: adversarial autoencoders[49], generative adversarial networks (GANs)[50], deep convolutional GANs (DCGANs)[51], and so on. While comprehensive exploration of these is beyond the scope of this manuscript, we note that many of these models do still suffer from the fact that the discovered lower dimensional representations are not explicitly crafted to be task-relevant[52]. In fact, we can demonstrate that an implementation of a GAN modified to allow selection of specific categories of reconstruction (a conditional GAN, or cGAN[53]), fails quite miserably when trained only 2% of the MNIST or Fashion MNIST datasets (see **Supplementary Material S4** and **Figure S3**). These considerations led to the development of InfoGAN[9], an unsupervised learning technique which modifies a generative adversarial network (GAN) in order to learn interpretable, low-dimensional representations. InfoGAN accomplishes this task by maximizing mutual information between noise in the GAN network and observations. Yet despite the tremendous success of InfoGAN[9], it is highly disadvantaged for the limited (sample-poor) data type targeted here. Specifically, InfoGAN's success has been demonstrated only on large-scale training datasets consisting of tens of thousands of training images. Further, exploring and characterizing latent spaces in GANs in general is highly nontrivial[54,55]; for these reasons, GANs generally do not accomplish the goal targeted by the TRACE network.

Attempts to mitigate the curse of dimensionality in fMRI datasets by pooling data across subjects to create larger training sets have of course been established to try to mitigate this significant challenge, including the shared response model[4], hyperalignment[56–58], and more recently decoder + autoencoder approaches[59]. However, while these can pool fMRI data to create more training exemplars, they do not explicitly seek subject-specific response patterns and instead presuppose that all subjects share a common response pattern.

In sum, although we do not benchmark TRACE against InfoGAN, hyperaligned data, or the expansive space of model variants, we argue that TRACE's utility is not only in its ability to distill task-relevant, low-dimensional representations, but also to do so in exemplar-limited, biological-scale datasets such as those collected in human neuroimaging experiments within a single subject.

### 3.3 Limitations

One limitation of the present approach is that we (deliberately) made TRACE and other models extremely simple (as in, few layers), which could have limited their performance. We did not investigate whether TRACE-like architecture (addition of a classifier on the bottleneck layer) would similarly improve performance for more complex networks, or whether multi-layer perceptrons or convolutional neural network (CNN) classifiers would surpass the simple logistic regression classifiers used here. We also could have opted to make the models deeper, with many hidden layers, which might have resulted in benefits in classification or reconstruction. However, we reiterate that we selected a simple architecture to be able to best evaluate TRACE's advantages over a "plain vanilla" fully-connected autoencoder, as more complex architectures could obscure TRACE's advantages. Future work may wish to explore other possible TRACE-like architectures.

It is also worth mentioning that for the sake of consistency we kept all hyperparameters for all networks and datasets the same. However, during training TRACE on a new dataset, it is always possible to tune the hyperparameters (learning rate, batchsize, regularization, etc) in order to achieve better performance (e.g., better bottleneck classification accuracy). Future studies may also more comprehensively explore the impact of specific hyperparameter tuning choices on TRACE's behavior.

### 3.4 Implications & future directions

Our findings have potentially exciting implications for the discovery of both low-dimensional representations and representations in the original (and anatomically- and/or functionally-relevant, in the case of fMRI) input space. For example, if a study's goal is to induce canonical target patterns of neural activity for a particular object category with real-time decoded neurofeedback (DecNef[27,28,37]), one might wish to instead 'de-noise' the data by maximizing reconstruction classifier accuracy instead of reconstruction class specificity. In the fMRI dataset presented here, reconstruction classifier accuracy peaked at about d=200. It is possible that in other fMRI datasets, reconstruction classifier accuracy might peak at a non-maximal bottleneck dimensionality, in which case it could be used to select the best dimensionality for the task at hand. Alternatively, one could choose to select optimal bottleneck dimensionality based on when reconstruction class specificity or classifier accuracy exceeds the analogous metric calculated directly from the raw input data. Here we showed that TRACE either exceeds these benchmarks sooner than other models, or does so even when other models do not. Thus, the process for selecting the best bottleneck dimensionality can flexibly adapt to an experimenter's goals, and future research seeking to use TRACE to extract neural patterns for use with DecNef should explore how different bottleneck dimensionalities impact the success of the neurofeedback process.

Regardless of the method one uses to select bottleneck dimensionality, it seems likely that TRACE can remove task-irrelevant information in a way that is useful for DecNef. To demonstrate this possibility, we did one final exploratory test. Recall that the fMRI dataset used in this study is in part overlapping with the dataset used by Taschereau-Dumouchel and colleagues[37], and as such we can directly compare their binary ("cat" versus "everything that is not a cat") decoding accuracy with the decoding accuracy we achieved on TRACE's reconstructions. To translate the reconstruction classifier accuracy we achieved to a binary scale, we counted a prediction to be correct if the correct class was in the top 20 (out of 40) of predicted classes from our one-versus-all classifier (with chance classification accuracy at 2.5%). Taschereau-Dumouchel and colleagues[37] observed binary logistic regression classification accuracies of 71.7% on average within-subject (~1 hour of fMRI data per person). (Relying on hyperalignment[56] to pool their 30 subjects and subsequently train such classifiers, they observed mean 82.4% using a 30-subject concatenated dataset). When we trained logistic regression classifiers on each individual subject (i.e., no hyperalignment) – some of whom are actually the original subjects in that former study – and translated the classification accuracies as described to be on the same scale as binary classification, we achieved the equivalent of 94.4% binary accuracy at bottleneck dimensionality d=30 (where reconstruction class specificity was maximized). Thus, TRACE facilitates distillation of class-specific representations in native space that are superior to the original representations themselves for this purpose.

Another interesting future possibility would be to investigate the extent to which TRACE excels over other methods as a function of neuroanatomical area – for the purposes of DecNef or simply to investigate neural representations themselves. Here, we focused on object representations in high level visual cortex (VTC), but in theory one could ask how early in the visual processing pipeline we might find evidence that task relevance plays a meaningful role. In the fMRI dataset used here, the task was for subjects to identify the object category of the image, and as a result the images were not standardized across lower level visual features such that object category did indeed covary with lower level visual properties such as color or spatial frequency (e.g., the background color of the 'dolphin' images is predominantly blue, whereas this is not the case for the 'key' images). Future studies may wish to use standardized images to investigate to what extent TRACE may assist in extraction of task-relevant representations versus low-level visual properties, depending on task and brain area; due to the limitations of the dataset used here for this first proof of concept, we leave these questions to future investigations.

Finally, we want to end with a brief note about TRACE's promise beyond fMRI and DecNef, at a broader scale including other types of biological-scale datasets relevant to human behavior, because fMRI data are not unique in their sparsity of sample size. For example, in biological image analysis or even human microbiome research, we are interested in learning generalizable and biologically informative "truths" about biological systems. However, precisely in the same way as a typical fMRI dataset, biological datasets are often limited in sample size. Given TRACE's success here, we hope that its capacity to discover task-relevant information *despite* undesirable ratios of samples to input-dimensions can help discover truths about other biological processes. Future studies should apply TRACE to other biological-scale datasets, with the goal of discovering representations relevant to those researchers and domains.

Earlier, it was discussed that the brain probably encodes a much lower dimensional manifold for object recognition than the thousands of voxels in fMRI data. Therefore, discovering lower dimensional representations that are in fact more task relevant can greatly help researchers to interrogate these lower dimensions. It is important to acknowledge that utilizing deep learning models such as TRACE comes with the caveat of a more difficult interpretation. Thus, full exploration of the latent, low-dimensional representations extracted by TRACE remains a subject for further investigations using available explainable artificial intelligence methods[60].

# 4. Methods

## 4.1 Methods overview

We proposed the "Task-Relevant Autoencoder via Classifier Enhancement" (TRACE) model and directly compared its behavior to that of a standard autoencoder (AE), a variational autoencoder (VAE), and principal component analysis (PCA) with equivalent internal architecture. TRACE is equivalent to the AE model in every respect, with the exception of a classifier branch which reads out directly from the AE network's bottleneck layer and which contributes to the overall loss function of the network. Likewise, VAE's architecture was chosen to be identical to AE except an additional KL-divergence loss function to make sure the features in the bottleneck followed a standard Gaussian distribution. We benchmarked TRACE and other models on the MNIST and Fashion MNIST datasets often used in machine learning model evaluation, and then applied both models to a previously-collected fMRI dataset to showcase TRACE's utility in extracting task-relevant low dimensional representations in a nonparametric regime.

We defined four output metrics to quantitatively evaluate the behavior of TRACE and other models, described below (**Methods Section 4.4**). These output metrics were evaluated as a function of bottleneck dimensionality -- essentially, how much each dataset can be compressed while retaining task-relevant information in the compressed representation (the bottleneck layer in all models). Details of datasets and model architectures are described in the next sections.

## 4.2 Datasets

### 4.2.1 MNIST dataset

The MNIST dataset consists of 60,000 handwritten 28x28 pixel grayscale images (i.e., image dimensionality of 784 pixels) of the 10 classes of digits (i.e., 0,1,...9 ) with their corresponding labels as the training set and 10,000 samples as the test set with a total of 7000 images per class (training and test set combined) that was collected by LeCun and colleagues[25], and is one of the most commonly used benchmarks to evaluate the performance of deep learning models.

### 4.2.2 Fashion MNIST dataset

The Fashion MNIST dataset[26] consists of 60,000 samples of 28x28 grayscale images (i.e., the dimensionality of 784 pixels) of 10 clothing categories (i.e., shirt, shoes, etc) with a total of 7000 images per class (training and test set combined)[26].

### 4.2.3 fMRI dataset

#### 4.2.3.1 Participants & task

The fMRI dataset used here was collected previously for several separate projects and was partially reported previously by Taschereau-Dumouchel and colleagues[37]. The dataset we used here contained 60 usable subjects' whole-brain data; one subject's data produced wildly unstable outcomes across all metrics tested here (see **Methods Section 4.4**) and so was excluded from the final analysis, and an additional 10 subjects had previously been collected with a different imaging sequence and are not included here. From this dataset, we therefore examined 59 healthy human participants who had viewed 3600 images from 40 categories (90 exemplars of each category, including 30 categories of animals [dogs, cats, snakes, etc.] and 10 categories of man-made objects [keys, chairs, airplanes, etc.]) while whole-brain BOLD responses were acquired. The images were shown for 0.98s each in mini-blocks (chunks) of 2, 3, 4, or 6 images of each category displayed sequentially, and the participants were asked to press a button whenever the category of images was changed. The data was collected from each participant in six runs with short breaks in approximately one hour while they were in the scanner. Each image was on screen for a duration of 0.98 s. The voxel activities of the ventral temporal cortex (VTC) were used as input to the model in order to find the latest features represented by VTC. See the methods reported by Taschereau-Dumouchel and colleagues[37] for further details, including information on informed consent and ethics approval for this previously-existing dataset.

#### 4.2.3.2 Image acquisition & preprocessing

The functional imaging data acquisition and preprocessing procedures for this existing dataset are previously described elsewhere[37], but are included here for completeness. Participants in the database were scanned at one of two 3T MRI scanners (Siemens Prisma and Verio) with a head coil. Whole brain functional data were acquired in 33 contiguous slices (TR = 2000 ms, TE =30 ms, voxel size = 3 × 3 × 3.5 mm 3, field-of-view = 192 x 192 mm, matrix size = 64 x 64, slice thickness = 3.5 mm, 0 mm slice gap, flip angle = 80 deg) oriented parallel to the AC-PC plane. High-resolution T1-weighted structural MR images (MP-RAGE sequence; 256 slices, TR = 2250 ms, TE = 3.06 ms, 5 voxel size = 1 × 1 × 1 mm 3, field-of-view= 256 x 256 mm, matrix size = 256 x 256, slice thickness = 1 mm, 0 mm slice gap, TI = 900 ms, flip angle = 9 deg.) were also obtained. Functional images were preprocessed using standard procedures, including slice timing correction, motion correction, realignment to the first functional image, and coregistration to the structural scan using SPM 12 (Statistical Parametric Mapping; www.fil.ion.ucl.ac.uk/spm). The anatomical mask of the target region of interest, VTC (fusiform, lingual/parahippocampal, and inferior temporal cortex), was selected using the Freesurfer (http://surfer.nmr.mgh.harvard.edu/) automated gray matter segmentation combining the ROI labels of *fusiform*, *inferior temporal*, *lingual*, and *parahippocampal*.

Voxels in this combined ventral temporal ROI were detrended and then deconvolved using the least-square separate approach[61,62]. This method creates an iterative general linear model for each trial individually, such that the design matrix contains one parameter modeling the current trial, and two parameters modeling all other trials in the design (e.g., even- and odd-numbered trials). This standardized method allows deconvolution of each trial in a rapid-event related design such as this one to obtain parameter estimates for each individual trial. This process results in a $N_{VTC,S}$ x 3600 images (90 exemplars of each category) timeseries, where $N_{VTC,S}$ refers to the number of voxels in the combined VTC ROI for each subject S; in this dataset $N_{VTC,S}$ ranged from ~40 to 55 MB in CSV format. This timeseries formed the dataset used in this project.

### 4.3 Models

#### 4.3.1 Standard autoencoder model (AE)

The standard AE model provides a baseline benchmark of model behavior in all datasets tested. This model is designed to be almost as simple as possible in order to maximally highlight any differences between TRACE and other models' behaviors. Thus, the standard AE architecture consists of input and output layers with dimensionality equal to the size of the datasets (784 for MNIST and Fashion MNIST, and 1726-3078 for fMRI), plus a single hidden layer with 1000 units in each of the encoding and decoding sections.

#### 4.3.1.1 Inputs

To facilitate comparisons across the three datasets tested and to facilitate faster model convergence, we standardized inputs by scaling their values: MNIST and Fashion MNIST inputs were scaled to take on values between 0 and 1 by dividing all values by 255, and fMRI inputs (parameter estimates from the single-trial deconvolution described above) were standardized by z-scoring because these parameter estimates can take on arbitrary real numbers without bound. To prevent leakage of any task-relevant information from test sets to the training set, the scaling factors were determined only on the training set, and then the same scaling parameters were used to scale the test sets.

#### 4.3.1.2 Activation functions

For the hidden layers, we used the hyperbolic tangent as the activation function in order to discover more complex nonlinear patterns in the data, as this function was reported previously to be more sensitive in capturing detailed and local information to represent the data with lower dimensions[63]. For the bottleneck layer of the network, we selected the linear activation function (i.e., no activation function) because in initial explorations we found that using a linear function resulted in discovering more task-relevant features in comparison to other activation functions, such as hyperbolic tangent, rectified linear unit (ReLu) and sigmoid, since features of the bottleneck layer under the linear function had higher accuracy in decoding the categories (e.g., in the case of MNIST dataset and with 2 dimensions in the bottleneck the accuracy increases by 15 percent; full data not shown). For the final decoding layer, a linear activation function was also chosen because the fMRI data are unitless and take arbitrary numbers and therefore are not confined to be within a specific boundary. Using a linear function allows the reconstructing layer to assign any value that minimizes the difference between output and input of the autoencoder, unlike most other typical activation functions (e.g., sigmoid, Tanh, etc) which usually have a confined output value hence those are not good choices to reconstruct fMRI data.

#### 4.3.1.3 Objective function

**Equation 1** shows the objective function for the standard autoencoder which was chosen as the mean square error (MSE):

$$L_R = \frac{1}{m \times n} \sum_{i=1}^{m} \sum_{j=1}^{n} (\hat{X}_{ij} - X_{ij})^2 \qquad (1)$$

Where $X$ is the input with $m$ samples and $n$ input-dimensions, and $\hat{X}$ is the reconstruction of the input.

To minimize the loss function, the Adam[64] implementation of stochastic gradient descent (SGD) was used and the learning rate was chosen to be 1e-4 and the batch size was set to 32. To prevent overfitting, the dropout technique was used for regularization with a dropout rate of 0.1.

#### 4.3.2 Task-Relevant Autoencoder via Classifier Enhancement (TRACE)

The Task-Relevant Autoencoder via Classifier Enhancement (TRACE) model is identical to the standard AE model with the exception that a logistic regression classifier was attached to the bottleneck (**Figure 1**). The activation function for this "decoder branch" of the network was the softmax function (also known as Boltzman distribution) which outputs a probability distribution for each class (e.g., Classes of 10 digits for MNIST). The loss function for this branch of the network was chosen to be categorical cross-entropy, i.e.:

$$L_{CE} = \frac{-1}{m} \sum_{i=1}^{m} \sum_{c=1}^{k} y_{ci} \, log(\hat{y}_{ci}) \qquad (2)$$

where $k$ denotes the number of the classes, $y$ is the label of observation, and $\hat{y}$ is the predicted label.

In the TRACE network, the final objective function is the summation of reconstruction loss and the categorical cross-entropy loss function (**Equations 1** & **2**), i.e.:

$$L_{TRACE} = L_R + L_{CE}$$
$$= \frac{1}{m \times n} \sum_{i=1}^{m} \sum_{j=1}^{n} (\hat{X}_{ij} - X_{ij})^2 - \frac{\alpha}{m} \sum_{i=1}^{m} \sum_{c=1}^{k} y_{ci} \, log(\hat{y}_{ci}) \qquad (3)$$

where $\alpha$, sets the weight for the classifier part of the loss function in order to control for its participation in updating the parameters.

### 4.3.3 Variational autoencoder (VAE)

The variational autoencoder used here is identical to AE in terms of the architecture and hyperparameters. The only difference is that unlike AE and TRACE we used a softmax function at the last layer in the case of MNIST and Fashion MNIST datasets. We did this because the network performed very poorly when we used a linear function (i.e., no activation function). The loss function was defined as follows:

$$L_{VAE} = L_R + L_{KL} \qquad (4)$$

Where $L_R$ is the reconstruction loss and $L_{KL}$, Kullback–Leibler (KL) divergence, is to calculate the distance between the encoder distribution $q(z|x)$ and a prior distribution $p(z)$ which was chosen to be a standard gaussian distribution. Therefore the $L_{KL}$ becomes

$$L_{KL} = -0.5 \sum_{z} [1 + log\sigma_z^2 - \mu_z^2 - \sigma_z^2] \qquad (5)$$

Where $\mu_z$ and $\sigma_z$ are mean and standard deviation of the latent space in the bottleneck.

### 4.3.4 Training and test data set size

The MNIST and Fashion MNIST datasets, each using 60,000 training and 10,000 test samples, are popular choices in benchmarking deep learning models. Having a high ratio of training samples for the number of data dimensions in these datasets (i.e., 60,000/748 = 76.5) makes them good candidates even for very deep neural networks. However, verifying the advantage of a model on these huge dataset is not necessarily applicable to dataset with much smaller sample to input-dimension size ratio (e.g., the real-world fMRI dataset we used here with approximate samples-to-input-dimensions ratio of ~1.5), as such a ratio might lead to overfitting and thus poor predictive capacity. To ensure that TRACE is powerful not only when the samples-to-input-dimensions ratio is large but also when available data is much sparser (such as in biological datasets that often suffer from small sample size), we truncated the MNIST and fashion MNIST datasets to explore how TRACE behaves under increasing data truncation.

To accomplish this, we trained all models at 10, 30, 50, 70, 90, 95, and 98 percent of data truncation at the optimal bottleneck dimensionality (i.e., d=2) for both MNIST and Fashion MNIST. Reducing the number of training samples to 95 and 98 percent decreases the samples-to-input-dimensions ratio such that it is approximately the same as the fMRI dataset also used here. The training set at each level of data truncation was the same for all models. To prevent overfitting caused by the limited sample size in the truncation analysis, we adjusted the stopping rule to stop the training when there is no improvement in the optimization process for 20 epochs, with a maximum training of 300 epochs. We then evaluated the behavior of all four-outcome metrics under all levels of data truncation by using the conventional hold-out test set in the MNIST and Fashion MNIST (i.e., 10,000 sample test set) (**see Section 2.1.3**). In the case of the fMRI dataset, we performed the training on 2700 training samples and tested the trained models on 900 held-out test samples. At the extreme level of 98% data truncation in the MNIST and Fashion MNIST datasets, the sample size is reduced to 1200 exemplars with each exemplar having 784 input-dimensions. Therefore, the ratio of samples to input-dimensions is about 1.5 which is comparable to sample-to-input-dimension ratio of fMRI dataset (i.e., ~1.5).

### 4.3.4 Implementation details for models

We explored several different values for $\alpha$ as a hyperparameter (**Equation 3**), and selected 0.01 for all results shown here because this value gave us the lowest MSE loss while having minimal effect on the categorical cross-entropy loss in comparison to higher values $\alpha$. The total loss was also the lowest when $\alpha$ was set to 0.01 (see **Supplementary Material S1**). For all datasets and all networks we used the same architecture and hyperparameter values. To implement these networks, we used the Keras functional API[65] with TensorFlow[66] backend, using available GPUs in Google Colab Pro and 12.7 GB of RAM.

## 4.4 Outcome metrics

In order to explore what is the best low dimensional feature space that explains within class characteristics while preserving the ability of the network to reconstruct the input, we evaluated four metrics as a function of bottleneck dimensionality (2, 5, 10, 15, 20, 30, 50, 60, 100, 150, 200, 250, 500, 784 [the maximum size of the MNIST and Fashion MNIST datasets]] nodes) in all models and for all three datasets (MNIST, Fashion MNIST, and fMRI): (1) *reconstruction fidelity,* (2) *reconstruction classifier accuracy*, (3) *bottleneck classifier accuracy,* and (4) *reconstruction class specificity*, as described below and shown in cartoons in **Figure 7**. Collectively, these metrics provide a broad picture of how bottleneck dimensionality can balance reconstruction fidelity with class fidelity, i.e., balances exemplar reconstruction capacity on each trial and per image class (the 'gist' of a category) with extraction of task-relevant features in the compressed representation in the bottleneck.

Importantly, the metrics (2) and (5) can be benchmarked against the equivalent metric applied to the *input* (i.e., original data). Specifically, reconstruction classifier accuracy and reconstruction class specificity can be compared to the equivalent assessments for the original, raw input data as a benchmark of all models' ability to extract meaningfully less noisy representations and push them back out into the original input dimensionality, such that comparing these metrics between input and reconstruction serves as a meaningful measure of noise reduction accomplished via each model.

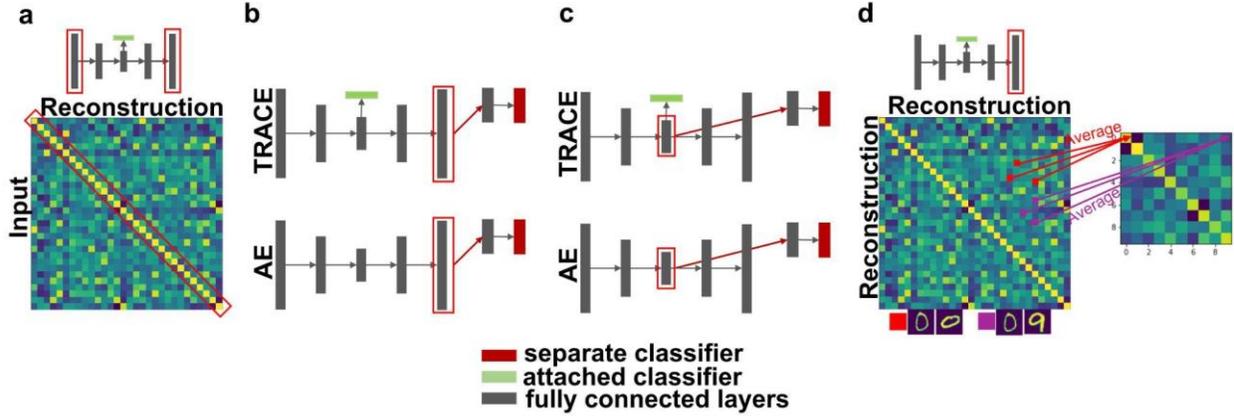

**Figure 7.** Graphical representation of the four quantitative outcome metrics. **(a)** Reconstruction fidelity (correlation between inputs and reconstructions; **Section 4.4.1)**, **(b)** reconstruction classifier accuracy (accuracy of a classifier trained on reconstructions; **Section 4.4.2**)), **(c)** bottleneck classifier accuracy (accuracy of a classifier trained on bottleneck features; **Section 4.4.3**), and **(d)** reconstruction class specificity (within- versus between-class separation; **Section 4.4.4**). Small cartoons of the TRACE architecture use red rectangle overlays to indicate which sections of the model architecture are being utilized for each outcome metric. In **(b)** and **(c)**, red-filled boxes indicate separate classifiers (i.e., trained separately after the main network has been trained), green-filled boxes indicate attached classifiers (i.e., which contribute to the main loss function for the network), and gray-filled boxes indicate fully-connected encoder and decoder layers.

### 4.4.1 Reconstruction fidelity

We quantified how well a model could reconstruct the input information with the average of all Pearson correlation coefficients between each input trial of the test set and the corresponding reconstruction of that sample (**Figure 7a**). We computed this correlation coefficient at all bottleneck dimensionalities tested. In the Results, below, this metric is referred to as *reconstruction fidelity* or $fidelity_R$.

$$fidelity_R = E(\rho_R) \qquad (6)$$

Where $\rho_R$ is the correlation between each input exemplar and its reconstruction, and $E$ denotes the expected value.

### 4.4.2 Reconstruction classifier accuracy

To quantify how well the reconstructed input represents a certain class, we used a separate logistic classifier (**Equation 7**) and trained it using reconstructed inputs for all dimensions in the bottleneck (**Figure 7b**). Using the same train/test folds as for training all models, we trained the data for 30 epochs for MNIST and Fashion MNIST and 300 epochs for fMRI.

$$L_{RCA} = \frac{1}{m}\sum_{i=1}^{m}\sum_{c=1}^{k} y_{ci}\, log(\hat{y}_{ci}) + \lambda \sum_{r=1}^{p} w_r^2 \qquad (7)$$

Where $L_{RCA}$ is the cross entropy loss for reconstructed input, and $\lambda$ is the regularization parameter and $w$ and $p$ are the weight matrices and the number of parameters of the classifier respectively.

### 4.4.3 Bottleneck classifier accuracy

We quantified the task-relevance of the features in the bottleneck via the accuracy of the logistic regression classifier with such bottleneck node activity as inputs (**Figure 7c**). For all models, this classifier is trained separately, *after* the training of models is finished. That is, this training occurs separately from the global loss function (which also includes an identical logistic regression classifier term, **Equations 2** & **3**) for all models.

To do this, we first extracted the bottleneck features after the training of all networks was complete, and then trained a completely separate logistic regression classifier to classify the category of the input image as it was now represented in the low-dimensional bottleneck of each model. The loss function for this separate classifier was identical to the one contributing to TRACE's global loss function (categorical cross-entropy), and the learning rate was set to 0.5e-4. To prevent overfitting of the classifier we used the L2 regularization technique. The final objective function for the separate logistic classifier is:

$$L_{BCA} = \frac{1}{m}\sum_{i=1}^{m}\sum_{c=1}^{k} y_{ci} \log(\hat{y}_{ci}) + \lambda \sum_{b=1}^{q} w_b^2 \qquad (8)$$

where $w$ and $q$ are the weight matrices and the number of parameters of the classifier respectively. The hyperparameter $\lambda$ was set to 0.007 which was manually tuned to maximize the classification accuracy.

### 4.4.4 Reconstruction class specificity

Another measure of the task-relevancy of the reconstructed information is the degree of similarity of representations within a class versus between classes. To this end, we computed the average within-class Pearson correlation of exemplars within each category from the reconstructions by extracting the pairwise correlation of all pairs of within-class exemplars for each category and finding the average of the correlation of them (we excluded the self-correlation of each trial in this calculation since it will always be 1). We then computed the average pairwise Pearson correlation across all pairs of trials from different classes. Graphically, this can be thought of as building an Nclass x Nclass similarity matrix (10x10 for the MNIST and Fashion MNIST data, 40x40 for the fMRI data), in which the diagonal represents the average pairwise within-class similarity (again, excluding self-similarity for a given exemplar) and the off-diagonals (**Figure 7d**). The final measure of representation class specificity from reconstructed inputs is thus the average of the diagonal (within class) of this similarity matrix minus the average of the off-diagonal (between class) of this matrix, i.e.

$$RCS = E(\rho_{R,within}) - E(\rho_{R,between}) \qquad (9)$$

where $RCS$ is the class specificity in the reconstruction of the input, and $\rho_{R,within}$ and $\rho_{R,between}$ are the Pearson correlation matrices between trials within each class and between one class and all other classes based on the reconstructed inputs excluding the elements of the main diagonal of the correlation matrix (i.e., the correlation of trial by itself was excluded since it is always 1 and does not reflect a true with class correlation), and $\rho_{R,between}$ is the Pearson correlation matrix of trials of the reconstructed inputs in one class and trials in all other classes.

We explored how this measure of class specificity changes as the dimension of the bottleneck increases. The purpose of this quantitative analysis is to discover the optimal bottleneck dimensionality for each dataset that can best distinguish within class versus between class categories in the reconstructions (i.e., in the original input space). In **Results** this metric is referred to as *reconstruction class specificity* or *RCS*.

### 4.4.5 Benchmarks against original inputs

To quantify the reduction in noise and the success of task-relevant feature extraction, we benchmark the reconstructions from all models in two ways.

First, we examined the classification accuracy of a simple logistic regression classifier applied to the input data in comparison to the accuracy of an identical classifier applied to the reconstructions (**Methods Section 4.4.2**). That is, if a representation has been successfully de-noised through the compression (and task-relevant feature extraction, in the case of TRACE), then the reduction in task-irrelevant noise should be apparent in the superior classification accuracy of a logistic regression classifier. Thus, we trained logistic regression classifiers on the input space as well as the reconstruction (**Methods Section 4.4.2**) at each bottleneck dimensionality, and reported the accuracy (**Equation 7**).

Second, a final test of the ability of models to extract task-relevant representations can be quantified via comparing the reconstruction class specificity (**Methods Section 4.4.4**) against input class specificity, calculated equivalently to reconstruction class specificity (**Equation 9**).

# Data/code availability statement

The data for this project are available from the corresponding authors upon reasonable request. The code implementing all models, including outcome metrics, is available at [GITHUB LINK AVAILABLE ON ACCEPTANCE].

# Author contributions

**Seyedmehdi Orouji:** Conceptualization, analysis, methodology, project administration, software, validation, visualization, writing -- original draft, writing -- review & editing. **Vincent Taschereau-Dumouchel, Aurelio Cortese, Brian Odegaard, Cody Cushing, & Mouslim Cherkaoui:** Data acquisition and preprocessing, validation, visualization, writing – review & editing. **Mitsuo Kawato & Hakwan Lau:** Conceptualization, validation, visualization, funding acquisition, writing – review & editing. **Megan A. K. Peters:** Conceptualization, funding acquisition, methodology, project administration, resources, supervision, validation, visualization, writing -- original draft, writing -- review & editing.

# Declaration of competing interests

None

# Acknowledgements

This work was supported in part by the Canadian Institute for Advanced Research Azrieli Global Scholars Program (M.A.K.P.), the *Fonds de Recherche du Québec - Santé* (V.T-C.), the Innovative Science and Technology Initiative for Security --ATLA (Grant Number JPJ004596; A.c.

and M.K.), and JST ERATO (grant number JPMJER1801; A.C. and M.K). Funding sources had no involvement in the design and methodology of the study.

# "Task-relevant autoencoding" enhances machine learning for human neuroscience

**Authors:** Seyedmehdi Orouji, Vincent Taschereau-Dumouchel, Aurelio Cortese, Brian Odegaard, Cody Cushing, Mouslim Cherkaoui, Mitsuo Kawato, Hakwan Lau, & Megan A. K. Peters

# Supplementary Material

## S1 Hyperparameter tuning

In order to tune the hyperparameter $\alpha$ to control for the contribution of the logistic classifier to the final objective function of TRACE, we chose different values for $\alpha$ (i.e., 0, 0.01, 0.1, 0.2, 0.5, 0.9, 1) for the bottleneck dimensionality of d=2. We chose $\alpha$ as 0.01 since it seemed it is the optimum point for the reconstruction and classification trade-off. Surprisingly, at $\alpha$=0.01 reconstruction loss was actually lower than $\alpha$=0 and the total loss was found to be almost the same as $\alpha$=0.

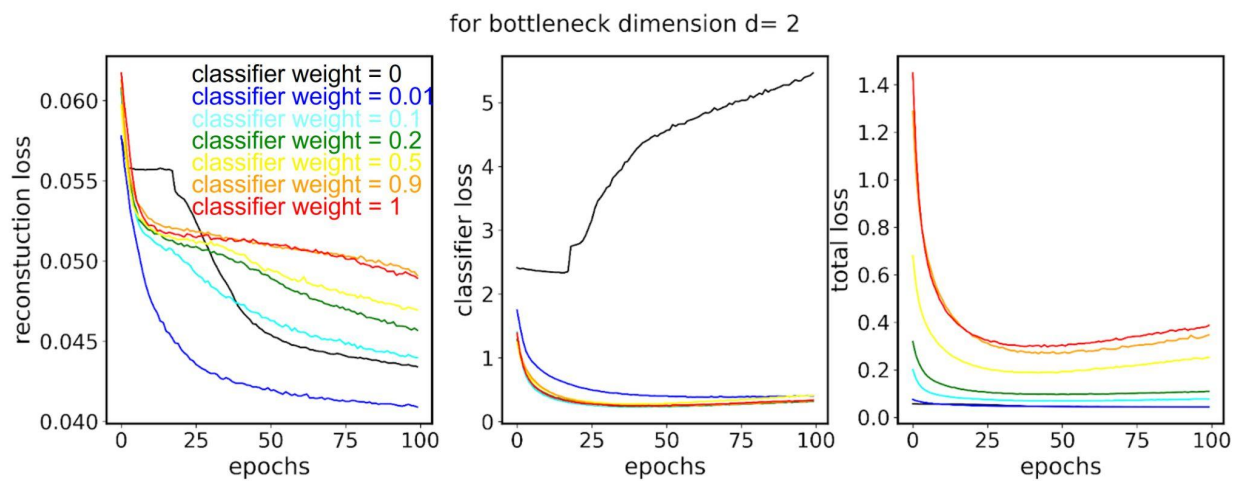

**Figure S1.** Loss functions for the MNIST dataset as a function of training epochs for bottleneck dimensionality d=2, for different levels of the hyperparameter $\alpha$. We selected $\alpha$=0.01 for all results presented in the main text because it showed the fastest convergence and achieved the smallest total loss (bottom panel) when training was complete.

## S2 Model fitting practical details

In order to meaningfully compare behaviors between TRACE and the other models (AE, VAE, and PCA), it is important to determine that both models can both be adequately fit to each of our datasets. Using available GPU processors in Google Colab Pro, it took about 61 minutes to fit the AE model to the MNIST training dataset (60,000 labeled samples), and about the same time for Fashion MNIST training dataset (60,000 labeled samples), and 21 minutes on average for each human subject in the fMRI dataset (3600 labeled samples of VTC) and for all 14 dimensionalities of bottleneck.

## S3 Comprehensive discussion of model performance

In the main text, we present and discuss the results from bottleneck dimensionalities between 2 - 150 for MNIST and Fashion MNIST, because at d>150 the quantitative metrics (see Methods) tend to asymptote. **Figure S2** presents the four metrics at bottleneck dimensions up to 784 as a comprehensive demonstration of this behavior. **Table S1** provides the bottleneck dimensionality at which each metric is maximized.

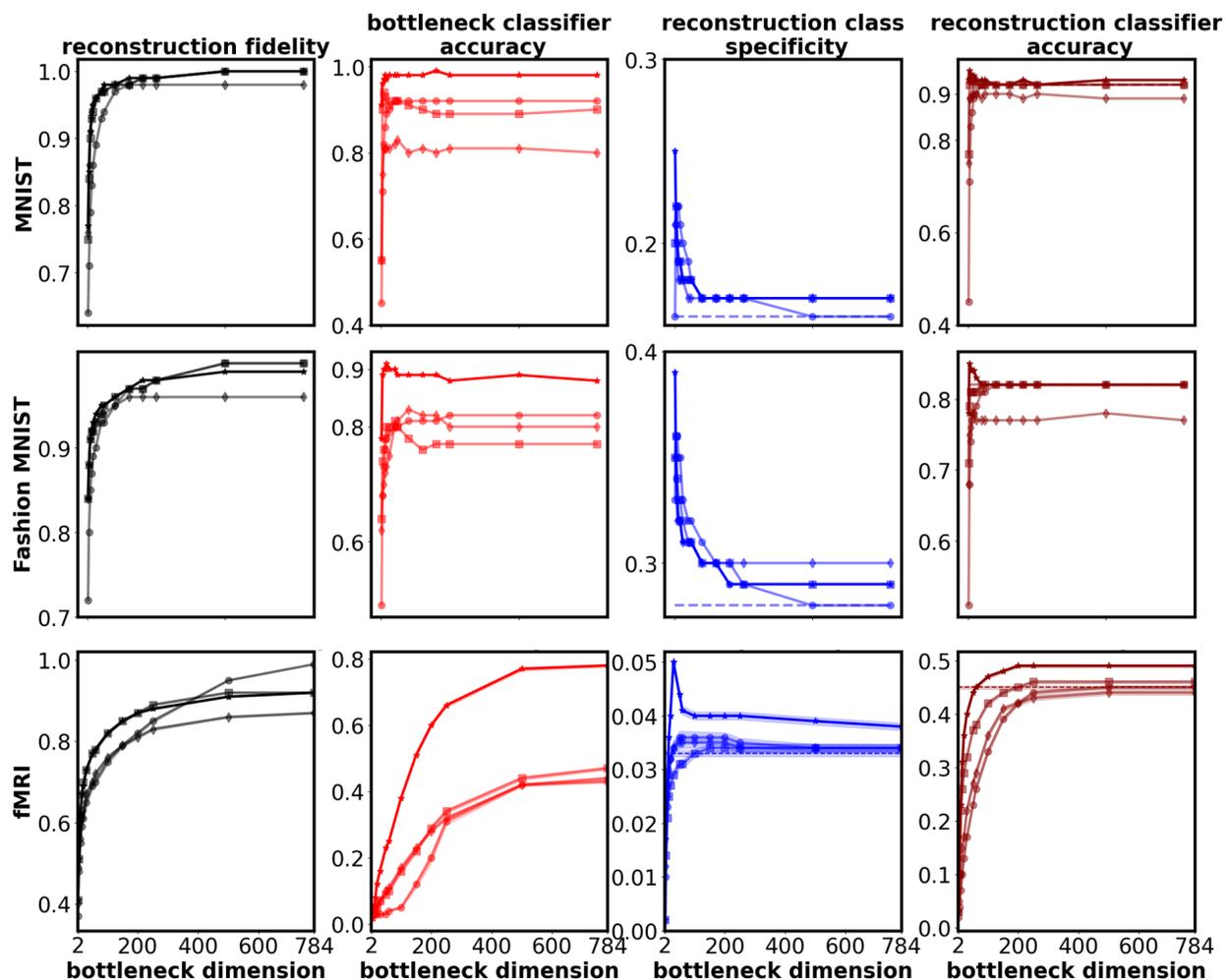

**Figure S2.** Quantitative comparison between TRACE and AE on the four outcome metrics, for all three datasets (MNIST, Fashion MNIST, and fMRI) for bottleneck dimensionalities between 2 and 784. As in **Figure 3**, the black, red, and blue, and dark red TRACEs show the reconstruction fidelity, bottleneck classifier accuracy, reconstruction class specificity, and reconstruction classifier accuracy, respectively. Here we partially replicate **Figure 6** from the main text to show comparisons directly between MNIST, Fashion MNIST, and fMRI data. As in the main text, dashed blue and dark red lines represent the input reconstruction class specificity and the input reconstruction classifier accuracy, respectively.

|   |   |   | MNIST | Fashion MNIST | fMRI |
|---|---|---|---|---|---|
| 1 | Reconstruction fidelity | TRACE | 500 | No peak | No peak |
|   |   | AE | 500 | 500 | No peak |
|   |   | VAE | No peak | No peak | No peak |
|   |   | PCA | 500 | 500 | No peak |
| 2 | Bottleneck classifier accuracy | TRACE | No peak | 20 | No peak |
|   |   | AE | 10 | 50 | No peak |
|   |   | VAE | 60 | 100 | No peak |
|   |   | PCA | No peak | No peak | No peak |
| 3 | Reconstruction class specificity | TRACE | 2 | 2 | 30 |
|   |   | AE | 5 | 5 | 50 |
|   |   | VAE | 5 | 2 | 50 |
|   |   | PCA | 10 | 5 | 50 |
| 4 | Reconstruction classifier accuracy | TRACE | 5 | 5 | No peak |
|   |   | AE | 10 | No peak | No peak |
|   |   | VAE | 15 | 15 | No peak |
|   |   | PCA | No peak | No peak | No peak |

**Table S1.** Bottleneck dimensionality at which each outcome metric is maximized.

|   |                                     |       | MNIST | Fashion MNIST | fMRI  |
|---|-------------------------------------|-------|-------|---------------|-------|
| 1 | **Reconstruction fidelity**         | **TRACE** | 0.77  | 0.84          | 0.73  |
|   |                                     | **AE**    | 0.75  | 0.84          | 0.73  |
|   |                                     | **VAE**   | 0.76  | 0.84          | 0.67  |
|   |                                     | **PCA**   | 0.64  | 0.72          | 0.65  |
| 2 | **Bottleneck classifier accuracy**  | **TRACE** | 0.91  | 0.78          | 0.16  |
|   |                                     | **AE**    | 0.55  | 0.64          | 0.07  |
|   |                                     | **VAE**   | 0.55  | 0.62          | 0.07  |
|   |                                     | **PCA**   | 0.45  | 0.49          | 0.03  |
| 3 | **Reconstruction class specificity**| **TRACE** | 0.25  | 0.39          | 0.05  |
|   |                                     | **AE**    | 0.2   | 0.35          | 0.029 |
|   |                                     | **VAE**   | 0.21  | 0.35          | 0.034 |
|   |                                     | **PCA**   | 0.16  | 0.33          | 0.034 |
| 4 | **Reconstruction classifier accuracy** | **TRACE** | 0.93  | 0.78          | 0.4   |
|   |                                     | **AE**    | 0.77  | 0.71          | 0.32  |
|   |                                     | **VAE**   | 0.75  | 0.68          | 0.22  |
|   |                                     | **PCA**   | 0.45  | 0.51          | 0.17  |

**Table S2**. Performance at bottleneck dimensionality d=2 for MNIST and Fashion MNIST, and d=30 for fMRI.

|  | **Reconstruction fidelity** | **Bottleneck classifier accuracy** | **Reconstruction class specificity** | **Reconstruction classifier accuracy** |
|---|---|---|---|---|
| **Main effect** | $F(3,147)=7927$, $p<1e-3$ | $F(3,147)=579$, $p<1e-3$ | $F(3,147)=1181$, $p<1e-3$ | $F(3,147)=2257$, $p<1e-3$ |
| **TRACE vs AE** | $t(49)=20.1$, $p<1e-3$ | $t(49)=11.03$, $p<1e-3$ | $t(49)=22.65$, $p<1e-3$ | $t(49)=23.8$, $p<1e-3$ |
| **TRACE vs VAE** | $t(49)=74.8$, $p<1e-3$ | $t(49)=34.26$, $p<1e-3$ | $t(49)=35.17$, $p<1e-3$ | $t(49)=53.3$, $p<1e-3$ |
| **TRACE vs PCA** | $t(49)=173.8$, $p<1e-3$ | $t(49)=35.07$, $p<1e-3$ | $t(49)=74.07$, $p<1e-3$ | $t(49)=70.4$, $p<1e-3$ |

**Table S3.** Results of ANOVAs and planned pairwise contrasts comparing TRACE to all other models at 98% data truncation for the MNIST and Fashion MNIST datasets, across all four metrics.

|  | **Reconstruction fidelity** | **Bottleneck classifier accuracy** | **Reconstruction class specificity** | **Reconstruction classifier accuracy** |
|---|---|---|---|---|
| **Main effect** | $F(3,171)=83.1$, $p<1e-3$ | $F(3,171)=356$, $p<1e-3$ | $F(3,171)=111$, $p<1e-3$ | $F(3,171)=433$, $p<1e-3$ |
| **TRACE vs AE** | $t(57)=-0.798$, $p=0.855$ | $t(57)=17.143$, $p<1e-3$ | $t(57)=18.822$, $p<1e-3$ | $t(57)=10.61$, $p<1e-3$ |
| **TRACE vs VAE** | $t(57)=7.682$, $p<1e-3$ | $t(57)=16.826$, $p<1e-3$ | $t(57)=11.836$, $p<1e-3$ | $t(57)=23.45$, $p<1e-3$ |
| **TRACE vs PCA** | $t(57)=9.825$, $p<1e-3$ | $t(57)=28.663$, $p<1e-3$ | $t(57)=10.055$, $p<1e-3$ | $t(57)=34.32$, $p<1e-3$ |

**Table S4.** Results of ANOVAs and planned pairwise contrasts comparing TRACE to all other models for the fMRI dataset, across all four metrics.

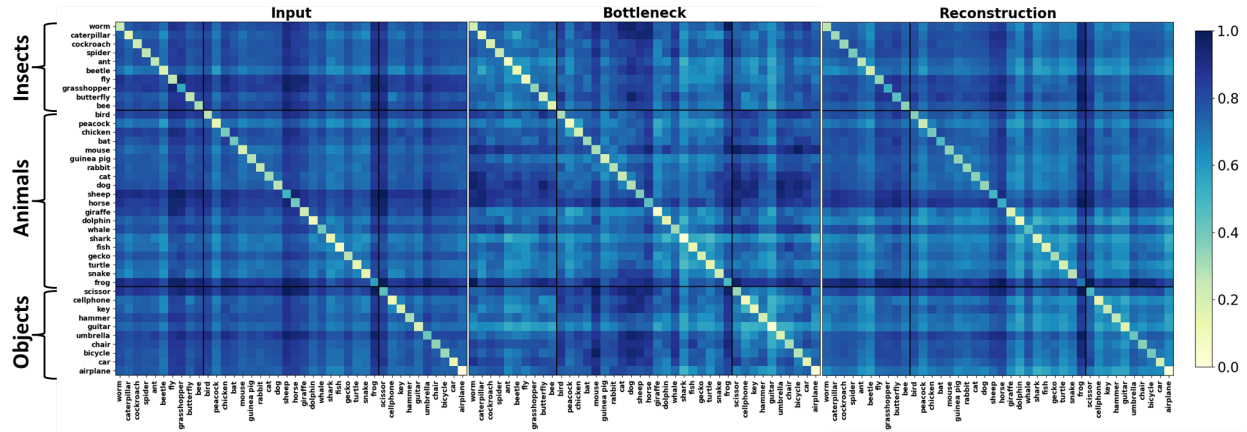

**Figure S3.** Representational dissimilarity matrices for input, bottleneck, and reconstruction in the fMRI dataset. Average Euclidean distances ($D(i,j) = \frac{1}{S \times M} \sum_{s=1}^{S} \sum_{m=1}^{M} d_{s,m}(i,j)$) where $S$ is the number of subjects, $M$ is the number of pairs between categories $i$ and $j$, and $d$ is the Euclidean distance between two pairs of exemplars; Euclidean distances are normalized within each plot such that the maximum distance is rescaled to 1 and the minimum distance rescaled to 0) between all pairs of objects (within- and between-classes), and across 58 subjects (one subject was excluded as an outlier). Clusters are visible in the bottleneck and reconstruction that are not visible in the input.

## S4 Comparison of reconstructions with other models

In addition to the three other benchmark models (i.e., AE, VAE, and PCA), we compared TRACE's reconstructions of MNIST and Fashion MNIST with reconstructions achieved with a conditional generative adversarial network (cGAN)[1]. As illustrated in **Figure S4**, the cGAN demonstrates strong performance when abundant data is available (e.g., 60,000 exemplars). However, the performance of cGAN significantly deteriorates in the face of data sparsity as depicted in **Figure S4** (e.g., 1200 exemplars). Therefore, using cGAN may not be a practical choice in the case of fMRI datasets which typically consist of only a few hundred exemplars.

We also benchmarked TRACE against a denoising autoencoder (DAE)[2]. However, we noticed that the DAE is primarily successful in eliminating the noise that we introduce to the input during training, and it fails to perform well when the structure of the noise is changed or unknown (which is always the case in fMRI datasets). Furthermore, DAE exhibited significant instability across some of the outcome metrics, leading us to opt against further exploration of this model or its inclusion in this manuscript.

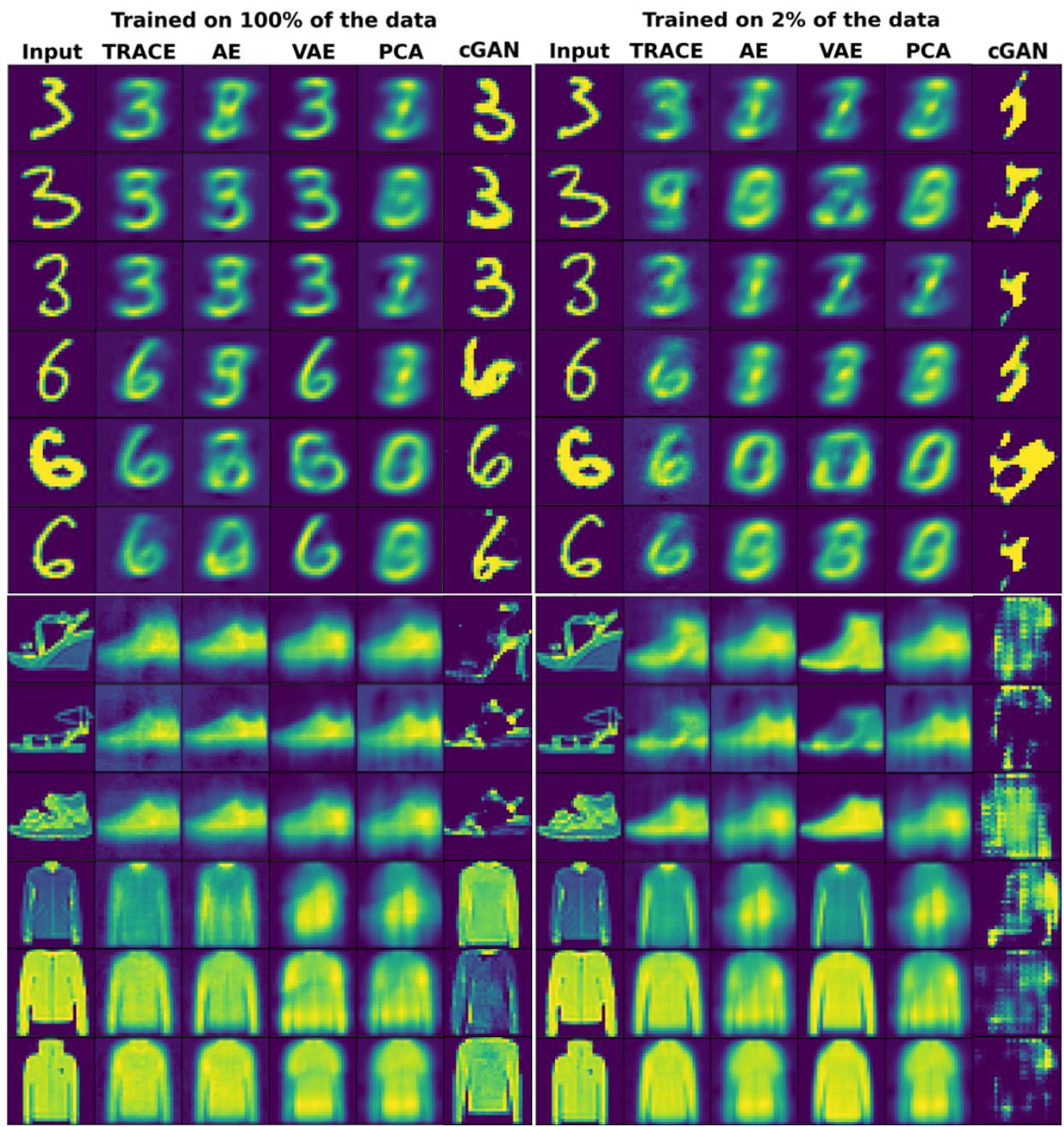

**Figure S4.** Reconstructions from all four models tested in the main text, plus a conditional generative adversarial network (cGAN). The cGAN performs quite well, according to visual inspection, when trained on 100% of the MNIST or Fashion MNIST datasets. However, when trained on only 2% of the data, the cGAN fails dramatically.